\documentclass[aps,prb,twocolumn,groupedaddress,noshowpacs,checklength,amsmath]{revtex4-1}

\usepackage{graphicx}
\usepackage{amssymb}
\usepackage{amsmath}
\usepackage{cancel}

\newcommand{\grad}{\mbox{\boldmath $\nabla$}}

\def\nhat{{\bf \hat n}}

\def\rcurs{{\mbox{$\resizebox{.16in}{.08in}{\includegraphics{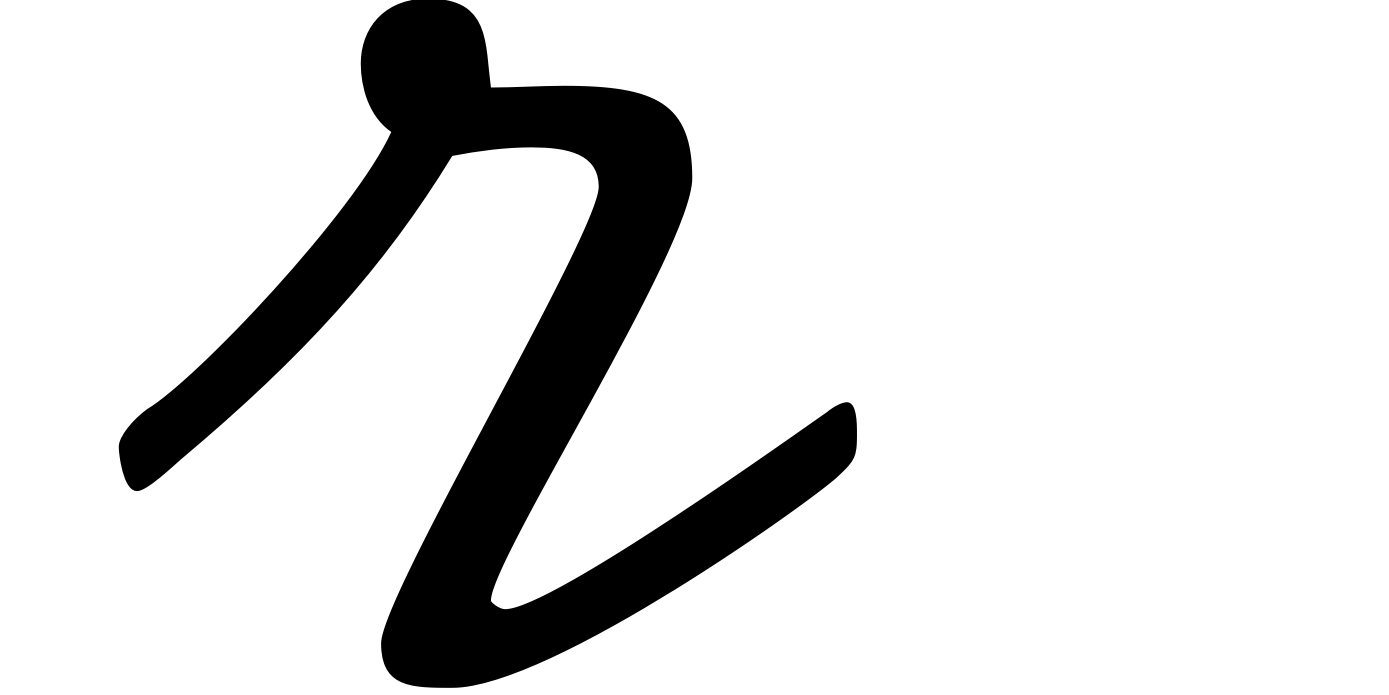}}$}}}
\def\brcurs{{\mbox{$\resizebox{.16in}{.08in}{\includegraphics{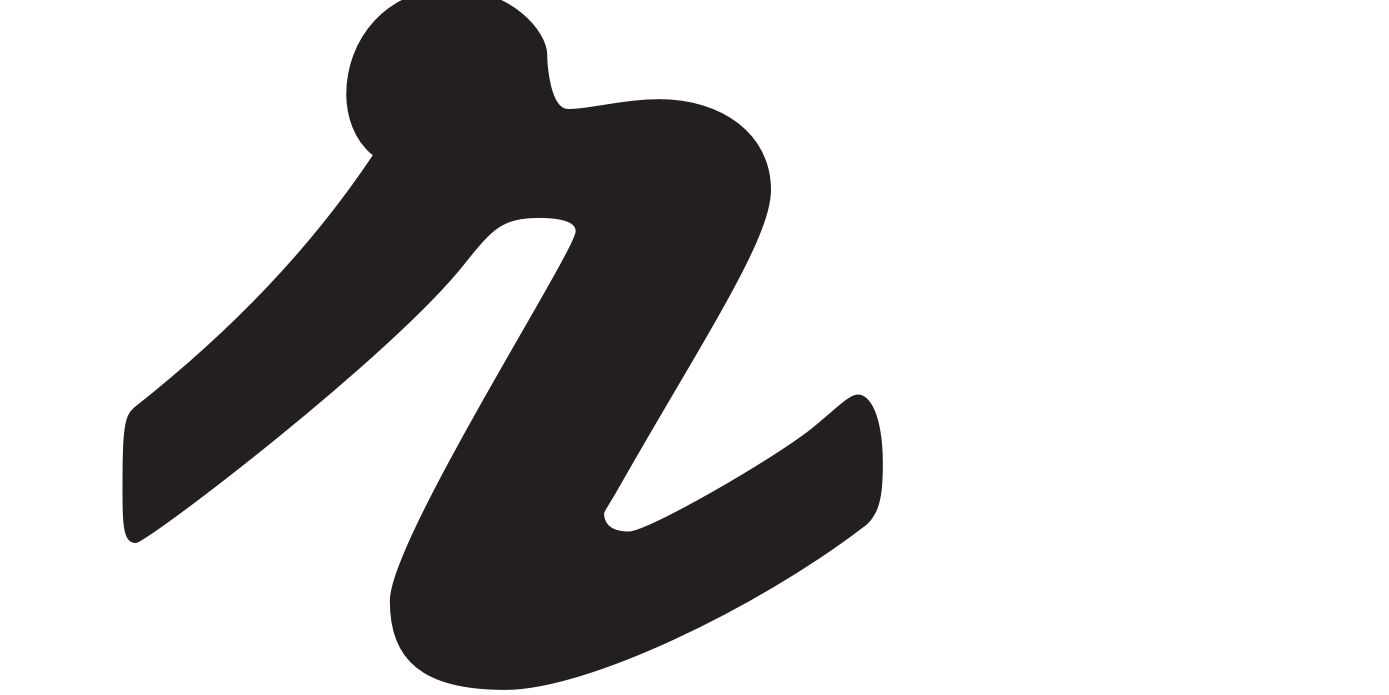}}$}}}
\def\hrcurs{{\mbox{$\hat \brcurs$}}}

\begin{document}

\title{What's the Use of Bound Charge?}

\author{David J.~Griffiths}\email[Electronic address: ]{griffith@reed.edu}
\affiliation{Department of Physics,
Reed College, Portland, Oregon  97202}
\author{V.~Hnizdo}\email[Electronic address: ]{hnizdo2044@gmail.com}
\affiliation{2044 Georgian Lane, Morgantown, WV  26508}

\begin{abstract}  
Bound charge is a useful construct for calculating the electrostatic field of polarized material, and it represents a perfectly genuine accumulation of charge.  But is such a material {\it in every respect} equivalent to a particular configuration of bound charge?  The answer is no, and the same goes for bound current and (in the time-dependent case) polarization current.
\end{abstract}

\maketitle

\section{Introduction}  

In introductory electrostatics we learn that the electric field of a polarized object (polarization ${\bf P} \equiv $ electric dipole moment per unit volume) is equivalent to the field produced by surface and volume ``bound" charges\cite{SUFC}
\begin{equation}
\sigma_b = {\bf P}\cdot \nhat,\quad \rho_b = -\grad\cdot {\bf P}\label{BC}
\end{equation}
where $\nhat$ is a unit vector perpendicular to the surface (pointing outward).  This is easy to understand: polarization results in perfectly genuine accumulations of charge,\cite{DG} differing from ``free" charge only in the sense that each electron is attached to a particular atom.\cite{PURCELL}  

But is polarized material {\it in every respect} equivalent to such a distribution of bound charge?  For example, is the electric {\it force} on a polarized object the same as it would be on $\rho_b$ and $\sigma_b$?  What about the {\it torque}?  And how about the force and torque {\it densities} within the material?  

The very notion of {\it density} (whether of force, torque, energy, or even mass, charge, and dipole moment) can be problematic.  After all, matter is composed of atoms, and on a microscopic scale these quantities fluctuate wildly in position and time.  We mean, however, their {\it macroscopic} averages over regions large enough to contain enormous numbers of atoms and yet small compared to the relevant dimensions of the object.

We will confine our attention to classical macroscopic {\it electromagnetic} forces.  Of course, the atoms in a solid or liquid are subject to all sorts of ``mechanical" forces (which may themselves be electromagnetic on a microscopic level), and they are governed by the laws of quantum mechanics.  But in this paper our purpose is to explore the role of bound charge, and to this end we adopt a radically simplified model:  We imagine a continuum of ideal (neutral) point dipoles, described by a specified function {\bf P}.  How this polarization came to be, and what ``mechanical" forces sustain it, we do not inquire.  (Imagine, in the static case, that they are simply {\it glued} in position.)  We are interested only in the electrical forces exerted on these dipoles by the (macroscopic) field {\bf E} (the {\it total} field, attributable both to the dipoles themselves and to any external sources).

If the question is ``How does a particular deformable medium respond to externally applied fields?" then one requires detailed information about the structure of the material, its elastic and dielectric properties, the pressure, the temperature, and so on.\cite{BOB}  Our question is much simpler: ``For a {\it stipulated} polarization, what is the {\it electromagnetic} force density, and in particular can it be calculated by replacing {\bf P} with the associated bound charge?"  We take it to be the ``correct" force density (as distinct from the force density associated with bound charge), but remember that it does {\it not} include the ``mechanical" stresses that would also be present in any real material.

In Section II we rehearse the standard derivation of the electrostatic potential of a polarized object, in terms of the bound charge.  We then apply the same reasoning to the force and torque on the object.  In Section III we do the same for static magnetization, and in Section IV we generalize to time-dependent configurations.  In Section V we compare our results with the Einstein-Laub force formula, and in Section VI we draw some lessons and conclusions.

\section{Fields and Forces for Polarized Matter}

Let's review how bound charge is first introduced:  The potential of an ideal dipole {\bf p} is
\begin{equation}
V({\bf r})=\frac{1}{4\pi\epsilon_0}\frac{{\bf p}\cdot \hrcurs}{\rcurs^2}
\end{equation}
(where $\brcurs \equiv {\bf r}-{\bf r}'$ is the vector from {\bf p}, at ${\bf r}'$, to the field point {\bf r}).
The potential of an object with polarization {\bf P} is therefore\cite{SUB}
\begin{equation}
V({\bf r})=\frac{1}{4\pi\epsilon_0}\int \frac{{\bf P}({\bf r}')\cdot \hrcurs}{\rcurs^2}\,d^3{\bf r}'.
\end{equation}
The standard integration by parts, using  
\begin{equation}
\grad'\left(\frac{1}{\rcurs}\right) = \frac{\hrcurs}{\rcurs^2},\label{DEL}
\end{equation}
turns this into
\begin{equation}
V=\frac{1}{4\pi\epsilon_0}\left[\int_{\cal V}\frac{(-\grad' \cdot {\bf P})}{\rcurs}\,d^3{\bf r}'+\int_{\cal S}\frac{{\bf P}\cdot \nhat}{\rcurs}da'\right],
\end{equation}
and we conclude that the potential of a polarized object is the same as that produced by the charge distribution $\rho_b$ and $\sigma_b$.

Next we ask  ``What is the {\it force}  on a piece of polarized material, in an electrostatic field {\bf E}?"  The force on an ideal dipole {\bf p} in an external field {\bf E} is\cite{DG2}
\begin{equation}
{\bf F} = ({\bf p}\cdot \grad){\bf E},\label{FoD}
\end{equation}
so the force on a chunk of polarized material is 
\begin{equation}
{\bf F} = \int ({\bf P}\cdot \grad){\bf E}\,d^3{\bf r}.\label{F1P}
\end{equation}
As before, we integrate by parts; the $i$th component is
\begin{eqnarray}
F_i &=& \sum_{j=1}^3\int P_j(\nabla_jE_i)\,d^3{\bf r}\nonumber\\
&=& \sum_{j=1}^3 \int\left[\nabla_j(P_jE_i)-E_i(\nabla_jP_j)\right]\,d^3{\bf r},\label{IBP1}
\end{eqnarray}
so\cite{AVE}
\begin{eqnarray}
{\bf F} &=& \int_{\cal V}(-\grad \cdot {\bf P})\,{\bf E}\,d^3{\bf r}+\oint_{\cal S}({\bf P}\cdot \nhat)\,\bar{\bf E}\,da\nonumber\\
&=&\int_{\cal V}\rho_b\,{\bf E}\,d^3{\bf r}+\oint_{\cal S}\sigma_b\,\bar{\bf E}\,da. \label{FP}
\end{eqnarray}
Thus the {\it total} force on the object is the same as it would be on $\rho_b$ and $\sigma_b$.

However, the force {\it densities} are {\it not} the same.  Equation \ref{F1P} says the force per unit volume is\cite{BOBBIO}
\begin{equation}
{\bf f} = ({\bf P} \cdot \grad){\bf E},
\end{equation}
whereas Eq.~\ref{FP} suggests
\begin{equation}
{\bf f}_b = \rho_b{\bf E}=-(\grad \cdot {\bf P}){\bf E} .\label{LOR}
\end{equation}
These two expressions are certainly not equivalent.  Imagine, for example,\cite{MAN} a ``bar electret" (a cylinder uniformly polarized along its axis); $\rho_b=0$ (the only bound charge resides on the two ends), so ${\bf f}_b={\bf 0}$, but if the field is nonuniform ${\bf f}\neq {\bf 0}$.  Bound charge incorrectly distributes the force, even though it gets the total force right.

This raises a surprisingly delicate question:  What do we {\it mean} by the ``force density" inside the medium?  Presumably we should (in the mind's eye) isolate an infinitesimal piece, of volume $v$, determine the force on it, and divided by $v$.  But this little piece carries {\it surface} bound charge in addition to its volume bound charge, and it's easy to see (reading Eqs.~\ref{F1P}-\ref{FP} in reverse) that the {\it total} force is precisely ${\bf f}v$.  Of course, in the bulk material the surface charge on $v$ is canceled by that on the adjacent inner surface of the surrounding medium---there is no {\it net} ``surface" charge within the substance---but if we're interested in the force on $v$ alone, its surface charge must not be ignored.  The force density ${\bf f}_b$ (Eq.~\ref{LOR}) is incomplete, because it does not include this contribution.

What about the {\it torque} on a polarized object, in a static electric field?  
The torque on an individual dipole is\cite{DG2}
\begin{equation}
{\bf N} = ({\bf p}\times {\bf E}) + [{\bf r}\times ({\bf p}\cdot \grad){\bf E}],\label{TOED}
\end{equation}
where {\bf r} is the vector to {\bf p} from whatever point we choose to calculate torques about.  The net torque on a polarized object, then, is
\begin{equation}
{\bf N} = \int({\bf P}\times{\bf E})\,d^3{\bf r} + \int[{\bf r}\times ({\bf P}\cdot\grad){\bf E}]\,d^3{\bf r} .\label{TQ}
\end{equation}
As always, we integrate by parts:
\begin{eqnarray}
\hskip-.2in N_i&=&\int\epsilon_{ijk}\left[P_jE_k+r_jP_l(\nabla_lE_k)\right]d^3{\bf r}\nonumber\\
&=&\int\epsilon_{ijk}[P_jE_k+\nabla_l(r_jP_lE_k)\nonumber\\
&&\quad\quad\quad-\,(\nabla_lr_j)P_lE_k- r_j(\nabla_lP_l)E_k)]d^3{\bf r}\nonumber\\
&=&\int\left\{\grad\cdot[{\bf P}({\bf r}\times {\bf E})_i] - (\grad\cdot {\bf P})({\bf r}\times{\bf E})_i\right\}d^3{\bf r},
\end{eqnarray}
(summation over repeated indices implied; $\nabla_lr_j=\delta_{lj}$), 
\begin{eqnarray}
{\bf N} &=& \oint ({\bf r}\times \bar{\bf E})({\bf P}\cdot \nhat)da - \int(\grad\cdot{\bf P})({\bf r}\times {\bf E})d^3{\bf r}\nonumber\\
&=&\oint_{\cal S} [{\bf r}\times (\sigma_b\bar{\bf E})]\,da + \int_{\cal V}[{\bf r}\times(\rho_b{\bf E})]\,d^3{\bf r}.\label{NT}
\end{eqnarray}
Again, the {\it total} torque on the object is the same as it would be on the bound charges.

However, Eq.~\ref{TQ} indicates that the torque {\it density} in the material is
\begin{equation}
{\bf n} = ({\bf P} \times {\bf E}) + ({\bf r}\times {\bf f}),
\end{equation}
whereas Eq.~\ref{NT} says it is
\begin{equation}
{\bf n}_b = {\bf r} \times \rho_b{\bf E}= ({\bf r}\times {\bf f}_b)\label{nL}.
\end{equation}
These expressions are not equivalent.  For example, if {\bf P} and {\bf E} are uniform, ${\bf n}_b = {\bf 0}$, whereas ${\bf n} =  ({\bf P} \times {\bf E})$---and surely there {\it is} a torque on the dipoles.  Once again, treating the medium as a configuration of bound charges gets the {\it total} right, but incorrectly assigns its distribution, because it ignores the role of ``internal" surface bound charge. 

\section{Magnetized Matter}

Now consider the magnetostatic analog: a chunk of magnetized material (${\bf M}\equiv$ magnetic dipole moment per unit volume).  The vector potential of an ideal dipole {\bf m} is
\begin{equation}
{\bf A} = \frac{\mu_0}{4\pi}\frac{{\bf m}\times \hrcurs}{\rcurs^2},
\end{equation}
so the potential of the magnetized object is
\begin{equation}
{\bf A}({\bf r}) = \frac{\mu_0}{4\pi}\int\frac{{\bf M}({\bf r}')\times \hrcurs}{\rcurs^2}\,d^3{\bf r}',
\end{equation}
and integration by parts (again using Eq.~\ref{DEL}) yields
\begin{equation}
{\bf A} = \frac{\mu_0}{4\pi}\left[\int_{\cal V}\frac{(\grad'\times {\bf M})}{\rcurs}\,d^3{\bf r}' + \oint_{\cal S} \frac{({\bf M}\times \nhat)}{\rcurs}\,da'\right].
\end{equation}
The two terms are identical to the potentials of (bound) volume and surface currents:\cite{SUFC}
\begin{equation}
{\bf J}_b = \grad\times {\bf M},\quad {\bf K}_b = {\bf M} \times \nhat.
\end{equation}
Once again, these are perfectly genuine currents, differing from free currents only in the sense that they are the collective effect of many tiny current loops---as in a relay race, no particular electron makes the entire trip.

But is magnetized material {\it in every respect} equivalent to the currents ${\bf J}_b$ and ${\bf K}_b$?  For example, are the forces on them the same?  The force on a magnetic dipole {\bf m} is\cite{FMD}
\begin{equation}
{\bf F} = {\bf m}\times(\grad\times {\bf B}) + ({\bf m}\cdot \grad){\bf B},\label{FOSD}
\end{equation}
so the force on a chunk of magnetized material is
\begin{equation}
{\bf F} = \int[{\bf M}\times(\grad\times {\bf B}) + ({\bf M}\cdot \grad){\bf B}]\,d^3{\bf r}.\label{FDM}
\end{equation}
From the vector identity
\begin{eqnarray}
\grad({\bf M}\cdot {\bf B}) &=& {\bf M}\times(\grad\times{\bf B}) + {\bf B}\times (\grad\times{\bf M})\nonumber\\
&&+\,({\bf M}\cdot \grad){\bf B}+({\bf B}\cdot\grad){\bf M}
\end{eqnarray}
it follows that
\begin{eqnarray}
{\bf F}&=&\int\Big[\grad({\bf M}\cdot{\bf B})-{\bf B}\times (\grad \times {\bf M})-({\bf B}\cdot \grad){\bf M}\Big]\,d^3{\bf r}\nonumber\\
&=&\int_{\cal V}({\bf J}_b\times{\bf B})\,d^3{\bf r} +{\bf G},
\end{eqnarray}
where 
\begin{equation}
G_i \equiv \int_{\cal V}\left[\nabla_i(M_jB_j) - B_j(\nabla_jM_i)\right]\,d^3{\bf r}.
\end{equation}
The second term in the integrand is $\nabla_j(B_jM_i)-M_i(\nabla_jB_j)=\nabla_j(B_jM_i) $, so\cite{AVE}
 \begin{eqnarray}
 G_i &=& \int_{\cal V}\left[\nabla_i(M_jB_j)- \nabla_j(M_iB_j)\right]\,d^3{\bf r}\nonumber\\
 &=&\oint_{\cal S}\left[\hat n_i(M_j\bar B_j) -\hat n_j(M_i\bar B_j)\right]\,da\nonumber\\
 &=&\oint_{\cal S}\left[({\bf M}\times \nhat)\times\bar{\bf B}\right]_i\,da\nonumber\\
 &=&\oint_{\cal S}\left[{\bf K}_b\times\bar{\bf B}\right]_i\,da.
 \end{eqnarray}
 Thus 
 \begin{equation}
 {\bf F}=\int_{\cal V}({\bf J}_b\times{\bf B})\,d^3{\bf r} +\oint_{\cal S}({\bf K}_b\times\bar{\bf B})\,da,\label{FSSA}
 \end{equation}
and the {\it total} force on the object is indeed the same as it would be for the bound current distributions.

However, the force {\it densities} inside the medium are different: the force per unit volume on ${\bf J}_b$ would be
\begin{equation}
{\bf f}_b = (\grad \times {\bf M})\times {\bf B},
\end{equation}
whereas the force density (from Eq.~\ref{FDM}) is
\begin{equation}
{\bf f} = {\bf M}\times(\grad\times {\bf B}) + ({\bf M}\cdot \grad){\bf B}.
\end{equation}

The torque on a magnetic dipole {\bf m} in a magnetostatic field {\bf B} is\cite{TMD}
\begin{equation}
{\bf N} = ({\bf m}\times{\bf B}) + {\bf r}\times[{\bf m}\times(\grad\times {\bf B}) + ({\bf m}\cdot \grad){\bf B}];
\end{equation}
the torque on a magnetized object is therefore
\begin{eqnarray}
\hskip-.4in{\bf N}&=& \int\Big\{({\bf M}\times{\bf B}) \nonumber\\
&&\quad\quad+\, {\bf r}\times[{\bf M}\times(\grad\times {\bf B}) + ({\bf M}\cdot \grad){\bf B}]\Big\}d^3{\bf r}.\label{TMTD}
\end{eqnarray}
Using the identity
\begin{equation}
\epsilon_{pqr}\epsilon_{pst} = \delta_{qs}\delta_{rt}-\delta_{qt}\delta_{rs},
\end{equation}
\begin{eqnarray}
N_i&=&\int\epsilon_{ijk}\Big\{M_jB_k +r_j[(\delta_{kn}\delta_{lp}-\delta_{kp}\delta_{ln})M_l(\nabla_nB_p)\nonumber\\&&\quad\quad+\,M_l(\nabla_lB_k)]\Big\}d^3{\bf r}\nonumber\\
&=&\int\epsilon_{ijk}\left[M_jB_k+r_jM_l(\nabla_kB_l)\right]d^3{\bf r}\label{SAAD}\\
&=&\int\epsilon_{ijk}\left[M_jB_k+\nabla_k(r_jM_lB_l)-r_jB_l(\nabla_kM_l)\right]d^3{\bf r}.\nonumber
\end{eqnarray}
Subtracting and adding
\begin{eqnarray}
\nabla_l(r_jM_kB_l) &=& (\nabla_lr_j)M_kB_l + r_j(\nabla_lM_k)B_l+ r_jM_k(\nabla_lB_l)\nonumber\\
&=&M_kB_j+r_jB_l(\nabla_lM_k)
\end{eqnarray}
to the expression in square brackets (last line of Eq.~\ref{SAAD}), we find
\begin{eqnarray}
\hskip-.2in N_i&=&\int\epsilon_{ijk}\Big\{\left[\nabla_k(r_jM_lB_l)-\nabla_l(r_jM_kB_l)\right]\nonumber\\
&&\quad\quad\quad +\,r_jB_l\left[(\nabla_lM_k)-(\nabla_kM_l)\right]\Big\}d^3{\bf r}.
\end{eqnarray}
We are now set up to integrate by parts, using 
\begin{equation}
\int_{\cal V}(\nabla_kQ)\,d^3{\bf r} = \oint_{\cal S}Q\,\hat n_k\,da,
\end{equation}
where the function $Q$ may carry one or more indices.  Thus
\begin{eqnarray}
N_i &=& \oint_{\cal S}\epsilon_{ijk}r_j\bar B_l\left[(M_l\,\hat n_k)-(M_k\,\hat n_l)\right]da\nonumber\\
&& +\,\int_{\cal V}\epsilon_{ijk}r_jB_l\left[(\nabla_lM_k)-(\nabla_kM_l)\right]d^3{\bf r},
\end{eqnarray}
and so
\begin{equation}
{\bf N} = \oint_{\cal S}\left[{\bf r}\times ({\bf K}_b\times \bar{\bf B})\right]da
+\int_{\cal V}\left[{\bf r}\times({\bf J}_b\times {\bf B})\right]d^3{\bf r}.\label{BCTD}
\end{equation}
Once again, the bound currents get the {\it total} torque right, but whereas the torque {\it density} (from Eq.~\ref{TMTD}) is 
\begin{equation}
{\bf n}=({\bf M}\times{\bf B})+ ({\bf r}\times{\bf f}),
\end{equation}
the bound currents suggest (Eq.~\ref{BCTD})
\begin{equation}
{\bf n}_b= {\bf r}\times({\bf J}_b \times {\bf B})=({\bf r}\times{\bf f}_b).
\end{equation}

\section{The Time-Dependent Case}

Consider an ideal (point) electric/magnetic dipole---its total charge is zero, but it carries an electric dipole moment ${\bf p}(t)$ and a magnetic dipole moment ${\bf m}(t)$.  Its position (${\bf r}'$) is fixed, but its dipole moments vary in magnitude and/or direction.  It produces scalar and vector potentials\cite{DDs}
\begin{eqnarray}
V({\bf r},t)& =& \frac{1}{4\pi\epsilon_0}\frac{\hrcurs}{\rcurs^2}\cdot\left[{\bf p}(t_r) + \frac{\rcurs}{c}\dot{\bf p}(t_r)\right],\label{VFDI}\\
{\bf A}({\bf r},t)& =& \frac{\mu_0}{4\pi}\Bigg\{\frac{\dot {\bf p}(t_r)}{\rcurs}\nonumber\\
&&\quad\quad-\, \frac{\hrcurs}{\rcurs^2}\times\left[{\bf m}(t_r) + \frac{\rcurs}{c}\dot{\bf m}(t_r)\right]\Bigg\},\label{AFDI}
\end{eqnarray}
where the dots denote time derivatives, and the sources are evaluated at the retarded time 
\begin{equation}
t_r = t-\frac{\rcurs}{c}.\label{RTT}
\end{equation}

The potentials of an object with time-dependent polarization and magnetization are therefore\cite{POSF}
\begin{eqnarray}
V & =& \frac{1}{4\pi\epsilon_0}\int\frac{\hrcurs}{\rcurs^2}\cdot\left[{\bf P}({\bf r}',t_r) 
 +\frac{\rcurs}{c}\dot{\bf P}({\bf r}',t_r)\right] d^3{\bf r}',\nonumber\\
{\bf A}& =& \frac{\mu_0}{4\pi}\int\Bigg\{\frac{\dot {\bf P}({\bf r}',t_r)}{\rcurs}\label{RPOTS}\\
&&\quad\quad-\, \frac{\hrcurs}{\rcurs^2}\times\left[{\bf M}({\bf r}',t_r) + \frac{\rcurs}{c}\dot{\bf M}({\bf r}',t_r)\right]\Bigg\}d^3{\bf r}'.\nonumber
\end{eqnarray}
As always, we use Eq.~\ref{DEL}, and integrate by parts:
\begin{eqnarray}
V&=& \frac{1}{4\pi\epsilon_0}\Bigg\{\int\grad'\cdot \left[\frac{1}{\rcurs}\left({\bf P}
 +\frac{\rcurs}{c}\dot{\bf P}\right)\right] d^3{\bf r}',\nonumber\\
 &&\quad \quad-\,\int\frac{1}{\rcurs}\grad'\cdot\left[{\bf P} + \frac{\rcurs}{c}\dot{\bf P}\right]d^3{\bf r}'\Bigg\}.\label{SPOT}
 \end{eqnarray}
Note that $\grad'$ acts not only on the explicit ${\bf r}'$ dependence in ${\bf P}({\bf r}',t_r)$, but also the {\it implicit} ${\bf r}'$ dependence in $t_r$.  Thus
\begin{equation}
\grad'\cdot{\bf P} = \tilde\grad'\cdot {\bf P} + \dot{\bf P}\cdot \grad't_r,
\end{equation}
where $\tilde\grad'\cdot {\bf P}$ denotes the divergence with respect to the the first argument (the explicit ${\bf r}'$) only.  Now, from Eq.~\ref{RTT},
\begin{equation}
\grad't_r= -\frac{1}{c}\grad'\rcurs,
\end{equation}
and
\begin{eqnarray}
\grad'\rcurs &= &\grad'\sqrt{(x-x')^2+(y-y')^2+(z-z')^2} \nonumber\\
&=& -\hrcurs.
\end{eqnarray}
So
\begin{eqnarray}
\grad'\cdot\left[{\bf P} + \frac{\rcurs}{c}\dot{\bf P}\right] &=& \left[\tilde\grad'\cdot{\bf P} +\frac{\hrcurs}{c}\cdot\dot{\bf P}\right] \nonumber\\
&&\quad\quad-\, \frac{\hrcurs}{c}\cdot\dot{\bf P}+ \frac{\rcurs}{c}\grad'\cdot\dot{\bf P}\nonumber\\
&=&\tilde\grad'\cdot{\bf P} + \frac{\rcurs}{c}\grad'\cdot\dot{\bf P}.\label{DIVP}
\end{eqnarray}
Meanwhile, the first term in Eq.~\ref{SPOT} can be converted to a surface integral:
\begin{eqnarray}
V&=&\frac{1}{4\pi\epsilon_0}\Bigg\{\oint_{\cal S}\frac{{\bf P}\cdot \nhat}{\rcurs}\,da'+ \frac{1}{c}\int_{\cal V}\grad'\cdot \dot{\bf P}\,d^3{\bf r}'\nonumber\\
&&\quad+\,\int_{\cal V}\left[\frac{(-\tilde\grad'\cdot{\bf P})}{\rcurs} - \frac{1}{c}\grad'\cdot \dot{\bf P}\right]d^3{\bf r}'\Biggr\}.
\end{eqnarray}
The two $\dot{\bf P}$ terms cancel, and we are left with
\begin{equation}
\hskip-.05in V({\bf r},t) = \frac{1}{4\pi\epsilon_0}\left[\int_{\cal V}\frac{\rho_b({\bf r}', t_r)}{\rcurs}d^3{\bf r}' + \oint_{\cal S}\frac{\sigma_b({\bf r}', t_r)}{\rcurs}da'\right].
\end{equation}
The bound charges are unchanged (Eq.~\ref{BC}), though they are evaluated, now, at the appropriate retarded times.

Turning to the vector potential (Eq.~\ref{RPOTS})
\begin{eqnarray}
{\bf A}&=&\frac{\mu_0}{4\pi}\Bigg\{\int \frac{\dot{\bf P}}{\rcurs} d^3{\bf r}' - \int\grad'\times\left[\frac{1}{\rcurs}\left({\bf M} +\frac{\rcurs}{c}\dot{\bf M}\right)\right]d^3{\bf r}'\nonumber\\
&&\quad+\,\int\left[\frac{1}{\rcurs}\grad'\times \left({\bf M} +\frac{\rcurs}{c}\dot{\bf M}\right)\right]d^3{\bf r}'\Bigg\}.
\end{eqnarray}
Proceeding as before, 
\begin{equation}
\grad'\times\left[{\bf M} + \frac{\rcurs}{c}\dot{\bf M}\right] =\tilde\grad'\times{\bf M} + \frac{\rcurs}{c}\grad'\times\dot{\bf M},\label{CRLM}
\end{equation}
and
\begin{eqnarray}
{\bf A}&=&\frac{\mu_0}{4\pi}\Bigg\{\int \frac{\dot{\bf P}}{\rcurs} d^3{\bf r}' - \int\grad'\times\left(\frac{{\bf M}}{\rcurs}\right)d^3{\bf r}'\nonumber\\
&&\quad\quad +\,\int\frac{\tilde\grad'\times{\bf M}}{\rcurs}d^3{\bf r}'\Bigg\}\nonumber\\
&=&\frac{\mu_0}{4\pi}\Bigg\{\int_{\cal V} \frac{{\bf J}_b({\bf r}',t_r)+ {\bf J}_p({\bf r}',t_r)}{\rcurs}d^3{\bf r'}\nonumber\\
&&\quad +\,\oint_{\cal S}\frac{{\bf K}_b}{\rcurs}da'\Bigg\},
\end{eqnarray}
where
\begin{equation}
{\bf J}_p \equiv \frac{\partial {\bf P}}{\partial t}.\label{POC}
\end{equation}
Again, the bound currents are unchanged (though they must now be evaluated at the retarded times), but they are joined by the polarization current (Eq.~\ref{POC}).

Next we calculate the {\it force} on the polarized/magnetized object.  To begin with, we need the force on point dipoles (${\bf p}(t)$ and ${\bf m}(t)$), in the presence of time-dependent fields.  The Lorentz force law says
\begin{equation}
{\bf F} = \int[\rho {\bf E} + ({\bf J}\times {\bf B})]\,d^3{\bf r}.\label{LORF}
\end{equation}
The charge and current densities for point dipoles (at the origin) are\cite{DFSS}
\begin{eqnarray}
\rho({\bf r},t) &=& -({\bf p}\cdot\grad)\,\delta^3({\bf r}),\label{CDEN}\\
{\bf J}({\bf r},t)&=& \dot{\bf p}\,\delta^3({\bf r}) -({\bf m}\times\grad)\,\delta^3({\bf r}),\label{CURR}
\end{eqnarray}
so
\begin{eqnarray}
{\bf F} &=&\int\Big\{-[({\bf p}\cdot \grad)\delta^3({\bf r})]{\bf E} + [\dot{\bf p}\delta^3({\bf r})]\times {\bf B}\nonumber\\ 
&&\quad\quad-\, [({\bf m}\times \grad)\delta^3({\bf r})]\times {\bf B}\Big\}d^3{\bf r}\\
&=&({\bf p}\cdot \grad){\bf E} + (\dot{\bf p}\times {\bf B}) + {\bf m}\times (\grad\times{\bf B})+ ({\bf m}\cdot \grad){\bf B},\nonumber
\end{eqnarray}
where {\bf E} and {\bf B} are evaluated at the location of the dipoles.  Except for the addition of the $\dot{\bf p}$ term, the force on time-dependent dipoles in time-dependent fields is unchanged from the static case (Eqs.~\ref{FoD} and \ref{FOSD}).

The total force on a chunk of polarized/magnetized material is thus (integrating by parts as in Eq.~\ref{IBP1}, and going through steps similar to those leading from Eq.~\ref{FDM} to Eq.~\ref{FSSA})
\begin{eqnarray}
{\bf F} &=& \int\Big[({\bf P}\cdot \grad){\bf E} + (\dot{\bf P}\times {\bf B}) + {\bf M}\times (\grad\times{\bf B})\nonumber\\
&&\quad\quad+\, ({\bf M}\cdot \grad){\bf B}\Big]\,d^3{\bf r}\label{ELFL1}\\
&=&\int_{\cal V}\left[\rho_b\,{\bf E} + ({\bf J}_b+{\bf J}_p)\times {\bf B}\right]\,d^3{\bf r}\nonumber\\
&&\quad\quad +\, \oint_{\cal S}\left[\sigma_b\bar{\bf E} + ({\bf K}_b\times \bar{\bf B})\right]\,da.\label{LFLTD}
\end{eqnarray}
This is precisely the force acting on the bound charges/currents and the polarization current.  As always, the bound quantities get the {\it total} force right.  But the force {\it density} suggested by Eq.~\ref{LFLTD},
\begin{equation}
{\bf f}_b = (-\grad\cdot {\bf P}){\bf E} + (\dot{\bf P}\times {\bf B}) + (\grad\times{\bf M})\times{\bf B},\label{LLFL}
\end{equation}
is not at all the same as the actual force density (Eq.~\ref{ELFL1})
\begin{equation}
{\bf f} = ({\bf P}\cdot \grad){\bf E} + (\dot{\bf P}\times {\bf B}) + {\bf M}\times (\grad\times{\bf B})+ ({\bf M}\cdot \grad){\bf B}.\label{EEEE}
\end{equation}

The torque on a (time-dependent) electric/magnetic dipole is\cite{TDTD}
\begin{eqnarray}
&&{\bf N} = ({\bf p}\times {\bf E}) + ({\bf m}\times{\bf B}) + {\bf r} \times\label{EQ65}\\
&&\quad\left[({\bf p}\cdot\grad){\bf E} + (\dot{\bf p}\times {\bf B})+ {\bf m}\times(\grad\times{\bf B}) + ({\bf m}\cdot\grad){\bf B}\right].\nonumber
\end{eqnarray} 
The total torque on a piece of polarized material is therefore
\begin{eqnarray}
{\bf N} &= &\int\Big\{({\bf P}\times {\bf E}) + ({\bf M}\times{\bf B})\nonumber\\
&&+ \, {\bf r} \times
 \big[({\bf P}\cdot\grad){\bf E} + (\dot{\bf P}\times {\bf B}) \label{TOQQ}\\
 && \quad+\, {\bf M}\times(\grad\times{\bf B}) + ({\bf M}\cdot\grad){\bf B}\big]\Big\}\,d^3{\bf r},\nonumber
\end{eqnarray} 
or, integrating by parts as before (Eqs.~\ref{NT} and \ref{BCTD}):
\begin{eqnarray}
{\bf N} &=& \oint_{\cal S}{\bf r}\times \left[\sigma_b\bar{\bf E} +({\bf K}_b\times \bar{\bf B})\right]da\nonumber\\
&&+\,\int_{\cal V}{\bf r}\times\left[\rho_b{\bf E} + ({\bf J}_b+{\bf J}_p)\times {\bf B}\right]d^3{\bf r}.\label{QRST}
\end{eqnarray}
Equation \ref{TOQQ} says the torque density is
\begin{equation}
{\bf n} = ({\bf P}\times{\bf E}) + ({\bf M} \times {\bf B}) + {\bf r} \times {\bf f}\label{TRUE}
\end{equation} 
(where ${\bf f}$ is given by Eq.~\ref{EEEE})
but Eq.~\ref{QRST} suggests a different torque density
\begin{equation}
{\bf n}_b = {\bf r} \times {\bf f}_b
\end{equation}
(where ${\bf f}_b$ is given by Eq.~\ref{LLFL}).

\section{The Einstein-Laub Formula}

The fundamental force law in classical electrodynamics is
\begin{equation}
{\bf F} = q[{\bf E} + ({\bf v} \times {\bf B})], \quad {\rm or} \quad {\bf f} = \rho {\bf E} + ({\bf J}\times {\bf B})\label{LFL}
\end{equation}
(known universally as the ``Lorentz force law").
If you separate the charge and current into free and bound parts,
\begin{eqnarray}
\rho &=& \rho_f + \rho_b = \rho_f -\grad\cdot {\bf P},\nonumber\\
 {\bf J} &=&{\bf J}_f + {\bf J}_b + {\bf J}_p =  {\bf J}_f + (\grad \times {\bf M}) + \frac{\partial {\bf P}}{\partial t},\label{BCS}
\end{eqnarray}
and substitute this in, you get Eq.~\ref{LLFL} (including now any {\it free} charge/current terms):
\begin{eqnarray}
{\bf f}_{\rm L}&=& \rho_f {\bf E} + ({\bf J}_f\times{\bf B})\nonumber\\
&& -\,(\grad\cdot {\bf P}){\bf E} +(\grad\times {\bf M})\times {\bf B} + (\dot {\bf P}\times {\bf B}).\label{LLLL}
\end{eqnarray}
In the optics community Eq.~\ref{LLLL} is sometimes itself called the ``Lorentz force law" (that's why we use the subscript L).\cite{MAN}  This terminology is misleading.  As we have seen, the substitution (Eq.~\ref{BCS}) is incorrect when calculating force and torque densities, though it does (when combined, of course, with the appropriate surface terms) yield the right {\it total} force and torque on an object.  By contrast, Eq.~\ref{EEEE} treats the material as a collection of electric and magnetic dipoles, not as a distribution of bound charges and currents:
\begin{eqnarray}
{\bf f}&=&  \rho_f {\bf E} + ({\bf J}_f\times{\bf B})+({\bf P}\cdot \grad){\bf E} \nonumber\\
&&+\, (\dot{\bf P}\times {\bf B}) + ({\bf M}\cdot \grad){\bf B} + {\bf M}\times (\grad\times{\bf B}).\label{EINS}
\end{eqnarray}

The fact that their {\it integrals} are equal suggests that ${\bf f}_{\rm L}$ and ${\bf f}$ differ by a total derivative.  Indeed,
\begin{eqnarray}
{\bf f}-{\bf f}_{\rm L}&=&({\bf P}\cdot \grad){\bf E}+(\grad\cdot{\bf P}){\bf E} +({\bf M}\cdot \grad){\bf B}\nonumber\\
&&\quad\quad+\,{\bf M}\times (\grad\times{\bf B})-(\grad\times{\bf M})\times {\bf B}\nonumber\\
&=&\grad({\bf M}\cdot{\bf B})+[({\bf P}\cdot\grad){\bf E}+(\grad\cdot {\bf P}){\bf E}]\nonumber\\
&&\quad\quad -\,[({\bf B}\cdot\grad){\bf M}+(\grad\cdot {\bf B}){\bf M}].
\end{eqnarray}
Now
\begin{eqnarray}
[({\bf P}\cdot\grad){\bf E}+(\grad\cdot {\bf P}){\bf E}]_i&=&P_j (\nabla_j E_i) + (\nabla_jP_j)E_i\nonumber\\
&=& \nabla_j(P_jE_i),
\end{eqnarray}
(and similarly for {\bf M} and {\bf B}), so
\begin{equation}
({\bf f}-{\bf f}_{\rm L})_i = \nabla_i(M_jB_j) +  \nabla_j[P_jE_i- M_iB_j].
\end{equation}

There is a final twist to the story.  By ``force" we mean, of course, the rate of change of momentum.  But in special relativity the momentum of a system consists of two parts: ``overt" momentum associated with motion of the center-of-energy, and ``hidden" momentum,\cite{HIDM} associated with internally moving parts but {\it not} reflected in motion of the system as a whole.  Thus
\begin{equation}
{\bf p} = {\bf p}_o + {\bf p}_h.
\end{equation}
If we are only interested in the overt motion, we might introduce an ``overt" force,
\begin{equation}
{\bf F}_o \equiv \frac{d{\bf p}_o}{dt} = {\bf F} -\frac{d{\bf p}_h}{dt}.
\end{equation}

Now, the hidden momentum of a magnetic dipole in an electric field is\cite{WHM} 
\begin{equation}
{\bf p}_h = \frac{1}{c^2}({\bf m}\times {\bf E}),
\end{equation}
so the overt force density on magnetized material is 
\begin{equation}
{\bf f}_o = {\bf f} - \frac{1}{c^2}\frac{\partial({\bf M}\times {\bf E})}{\partial t}.
\end{equation}
Thus
\begin{eqnarray}
{\bf f}_o &=&  \rho_f {\bf E}+ ({\bf J}_f \times {\bf B}) +({\bf P}\cdot \grad){\bf E} + (\dot{\bf P}\times {\bf B}) \nonumber\\&&\quad\quad+\, {\bf M}\times (\grad\times{\bf B})+ ({\bf M}\cdot \grad){\bf B}\nonumber\\
&&\quad\quad-\,\frac{1}{c^2}(\dot{\bf M} \times {\bf E}) - \frac{1}{c^2}({\bf M}\times \dot{\bf E}).
\end{eqnarray}
This is {\it almost} the ``Einstein-Laub" force density,\cite{ELFD}
\begin{eqnarray}
{\bf f}_{\rm EL}&=&  \rho_f {\bf E}+ [{\bf J}_f \times (\mu_0{\bf H})]+({\bf P}\cdot \grad){\bf E} + \dot{\bf P}\times (\mu_0{\bf H})\nonumber\\
&&\quad\quad +\, ({\bf M}\cdot\grad)\mu_0{\bf H}-\frac{1}{c^2}\dot{\bf M}\times {\bf E}\nonumber\\
&=&\rho_f {\bf E}+ ({\bf J}_f \times {\bf B})+({\bf P}\cdot \grad){\bf E} + (\dot{\bf P}\times {\bf B})\nonumber\\
&&\quad\quad +\, ({\bf M}\cdot\grad){\bf B}-\mu_0\Big[({\bf J}_f\times {\bf M})+(\dot{\bf P}\times{\bf M})\nonumber\\
&&\quad\quad+\,({\bf M}\cdot\grad){\bf M} +\epsilon_0(\dot{\bf M}\times {\bf E})\Big].\label{ELFL}
\end{eqnarray}
In fact, using
\begin{eqnarray}
\grad\times{\bf B}&=& \mu_0{\bf J} + \frac{1}{c^2}\dot{\bf E}\nonumber\\
&=&\mu_0\left({\bf J}_f + \dot{\bf P} + \grad\times{\bf M} + \epsilon_0\dot{\bf E}\right),
\end{eqnarray}
we get
\begin{equation}
{\bf f}_{\rm EL}={\bf f}_o-\frac{\mu_0}{2}\grad(M^2).\label{EXTR}
\end{equation}
Since the ``extra" term ($-(\mu_0/2)\grad(M^2)$) is a pure gradient, it will not affect the total force on an object--but it does, of course, change the force {\it density}.\cite{ELDV} 

The same considerations apply to torque: the total angular momentum consists of two parts,
\begin{equation}
{\bf L} = {\bf L}_o+{\bf L}_h.
\end{equation}
The overt torque is
\begin{equation}
{\bf N}_o = \frac{d{\bf L}_o}{dt}=\frac{d{\bf L}}{dt} - \frac{d{\bf L}_h}{dt},
\end{equation}
where
\begin{equation}
{\bf L}_h = {\bf r}\times {\bf p}_h = \frac{1}{c^2}{\bf r}\times({\bf m}\times {\bf E})
\end{equation}
is the hidden angular momentum of the magnetic dipole.  Thus the overt torque density on polarizable/magnetizable material is (cf.~Eq.~\ref{TRUE})
\begin{equation}
{\bf n}_o=({\bf P}\times{\bf E}) + ({\bf M} \times {\bf B}) + {\bf r}\times {\bf f}_o.
\end{equation}
Meanwhile the Einstein-Laub torque density is 
\begin{eqnarray}
{\bf n}_{\rm EL} &=& ({\bf P}\times{\bf E}) + ({\bf M}\times{\bf B})+{\bf r}\times {\bf f}_{\rm EL}\nonumber\\
&=&{\bf n}_o-\frac{\mu_0}{2}{\bf r}\times (\grad M^2).\label{ELTD}
\end{eqnarray} 
Notice that ${\bf n}_o$ and ${\bf n}_{\rm EL}$ yield the same {\it total} (overt) torque on an object, though they describe rather different torque {\it densities}.

In recent years some authors\cite{MANS} have advocated the Einstein-Laub force law (Eq.~\ref{ELFL}), as a replacement for what they call the ``Lorentz" law (Eq.~\ref{LLLL}).  We agree that the latter is defective, but proponents of the former should be aware that they are only talking about the ``overt" part of the force density, and including an extra term (Eq.~\ref{EXTR}) of dubious provenance.  

\section{Conclusion} So, what {\it is} the use of bound charge (and bound current and polarization current)?  When is the substitution
\begin{eqnarray*}
\rho &=& \rho_f + \rho_b = \rho_f -\grad\cdot {\bf P},\nonumber\\
 {\bf J} &=&{\bf J}_f + {\bf J}_b + {\bf J}_p =  {\bf J}_f + (\grad \times {\bf M}) + \frac{\partial {\bf P}}{\partial t},
\end{eqnarray*}
(Eq.~\ref{BCS}) legitimate?  {\it Answer:} it's fine for calculating potentials and fields, and hence for use in Maxwell's equations.  It's OK when you are interested in {\it total} forces and torques.  But it does {\it not} yield the right force and torque {\it densities}---it distributes the force (over the object) incorrectly, even though it gets the total right.  There is nothing wrong with the Lorentz force law (Eq.~\ref{LFL}) itself.\cite{LVEL}  The problem, rather, is that the substitution Eq.~\ref{BCS} does not take proper account of the ``internal" surface bound charge and current.

\section*{Acknowledgment}
We thank Kirk McDonald and Masud Mansuripur for many useful discussions, and an anonymous referee for suggesting the references in endnote 4.

\appendix


\begin{thebibliography}{99}
\bibitem{SUFC} Surface bound charges (and currents) are implicitly included in the volume terms, as delta-function discontinuities at the boundary, but it is generally safer to handle them as separate contributions.
\bibitem{SUB} Actually, there is some subtlety to this, when applied to points inside the material (see Section 4.2.3 in ref.~2, or Section 2.6 of Bobbio, ref.~4).
\bibitem{FMD} See, for example, ref.~2, Eq.~6.3.   This formula is often written in the equivalent form ${\bf F} = \grad({\bf m}\cdot {\bf B})$, but that could be misleading in the present context, since {\bf m} is a fixed quantity, and is not to be differentiated.
\bibitem{TMD} See, for example, ref.~2, Sect.~6.1.2.  As in Eq.~\ref{TOED}, the first term is the torque about the center of the dipole, and second is $({\bf r}\times{\bf F})$, where {\bf F} is the force on the dipole.
\bibitem{WHM}  Hidden momentum resides in the moving parts of a system, in this case the magnetic dipoles.  There is no hidden momentum in the {\it electric} dipoles (though there may be hidden momentum in the source of the external magnetic field---which is not relevant here).
\end{thebibliography}
\end{document}